\documentstyle[aps,multicol,pre,epsf]{revtex}

\begin{document}

\draft
\title{Stabilizing Grafted Colloids in a Polymer Melt: \\
        Favorable Enthalpic Interactions\footnote[2]{Accepted for 
publication in Phys. Rev. E} }
\author{Itamar Borukhov\footnote{Current Address: 
Department of Chemistry and Biochemistry,
UCLA, 607 C. Young Dr. East, Los Angeles, CA 90095-1569, USA}
and Ludwik Leibler} 
\address{Unit\'e Mixte de Recherche CNRS/Elf Atochem (UMR 167), \\ 
95 rue Danton, B.P. 108, 92303 Levallois-Perret Cedex, France}

\date{\today}

\maketitle

\begin{abstract}
The interactions between spherical colloids covered with
end-grafted polymers (brushes) immersed in a polymer melt are
studied theoretically.
We show that attractive enthalpic interactions between the two
polymer species, characterized by a negative Flory parameter
($\chi<0$), can stabilize the colloidal dispersion. The stabilizing
mechanism is a result of a structural change from a dry brush,
where the melt chains do not penetrate deeply into the brush, to a
wet brush, where the free chains penetrate the brush and force the
grafted chains to extend into the melt.
\end{abstract}


\begin{multicols}{2}

\paragraph*{Introduction:}
Mixtures of colloids and polymers provide an important class of
advanced materials with a wide range of mechanical, adhesive and
optical properties. In many applications polymer chains are
end-grafted to the colloid surface forming what is usually
referred to as a {\em polymer brush}
\cite{Alexander,deGennes,parabolic}. When the grafted particles
are immersed in solution steric interactions acting between the
grafted chains stabilize the colloidal suspension against
attractive van der Waals interactions. Unfortunately, when the
embedding medium is a polymer melt this mechanism does not always
work. Due to subtle entropic effects the melt chains are expelled
from the grafted layer even if they are chemically identical to
the grafted chains
\cite{deGennes,Brown,WittenLeibler,Mourran,BrushMelt1,BrushMeltexp}
and the inter-colloidal interactions become attractive
\cite{BrushMelt2}. This attraction limits our ability to produce
new materials where the contrast between the colloidal and polymeric
properties might have advantageous implications.
The aim of this letter is to put forward a
simple mechanism of stabilizing the grafted colloids in the
surrounding polymer melt by means of a favorable enthalpic
attraction between the grafted and free chains. A number of such
systems exists, e.g., poly(methyl methacrylate) (PMMA) and
poly(vinyl chloride) (PVC), poly(styrene) (PS) and
poly(2,6-dimethyl-1,4-phenylene oxide) (PXE), etc.
 In practice, these specific interactions
have been used, for example, in the design of impact modifiers.
Here, we quantify this effect and provide simple criteria for
predicting the inter-colloidal interactions.

Colloidal systems are not the only example where enthalpic
interactions compete with entropy in determining the structure of
the material. Similar behavior can be also observed in mixtures of
homopolymers and diblock copolymers \cite{Semenov92}. Block
copolymer chains are formed from two (or more) different polymers
that are tethered together at a single point. For example, an AB
diblock copolymer consists of a sequence of A monomers followed by
a sequence of B monomers. Depending on the incompatibility of the
two blocks these systems have a tendency to segregate into A-rich
and B-rich mesophases \cite{diblocks}. The resulting phases
include lamellar structures of alternating A-rich and B-rich
stripes, hexagonal phases of cylindrical domains, cubic phases of
spherical domains, etc. Triblock copolymers (e.g. ABA or ABC)
exhibit an even larger variety of phases \cite{diblocks}. When
homopolymers are mixed with copolymers the mixture may exhibit
very interesting physical properties (e.g. mechanical, optical,
etc.) and can be used as, e.g. super tough materials or photonic
devices \cite{Thomas}. Many of these properties depend crucially
on the special meso-structure of the material and we would like to
be able to determine under what conditions homopolymers can be
added without destabilizing the structure.

\begin{figure}[tbh]
  \epsfxsize=0.4\linewidth
  \centerline{\hfill\vbox{ \epsffile{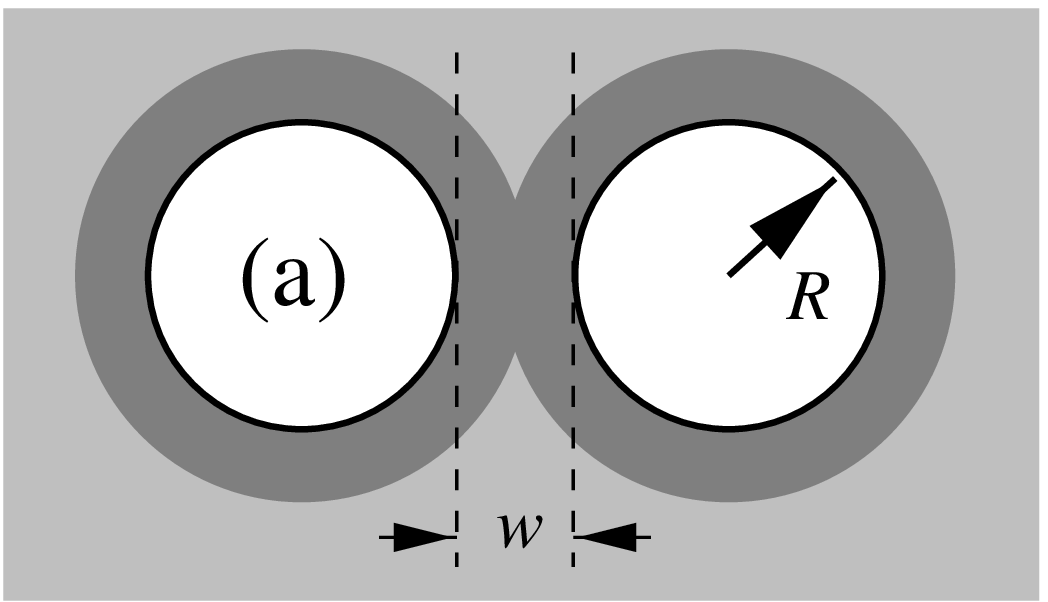}}
  \hfill
  \epsfxsize=0.4\linewidth
              \vbox{ \epsffile{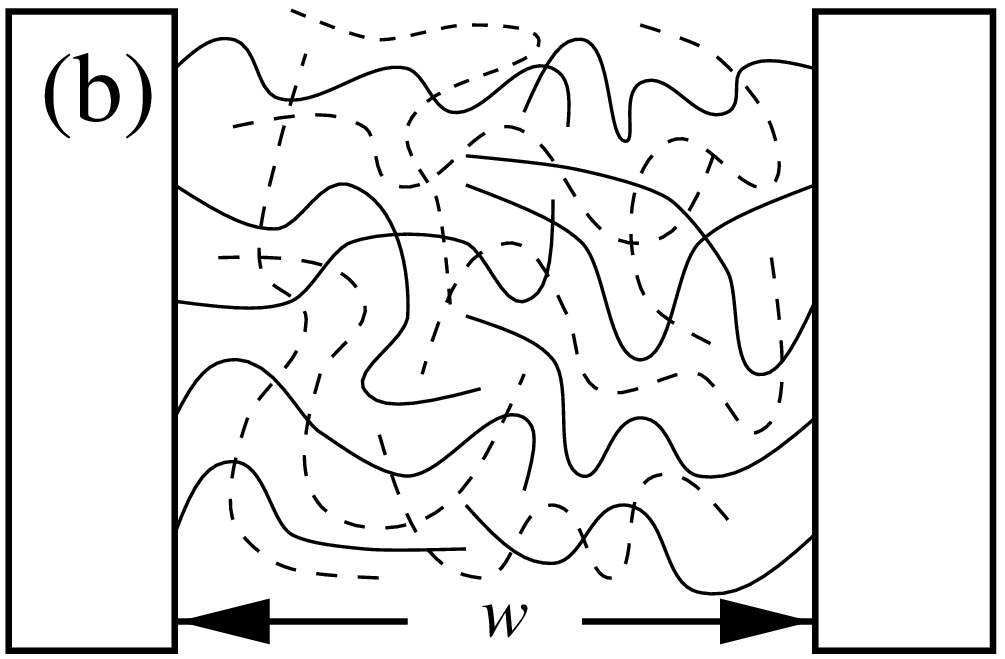}}\hfill}
\vspace{\baselineskip}
\caption{\protect Schematic views of 
(a) two grafted colloids immersed in a polymer melt and 
(b) two interacting flat surfaces in a polymer melt. 
The solid (dashed) lines are the grafted (free)
chains.}
\label{FigSchematic}
\end{figure}

Our model system consists of two grafted spheres immersed in a
polymer melt (Fig.~\ref{FigSchematic}a). Typically, the radii $R$
of the colloids is of the order of microns while the attractive
interactions appear at separations $w$ of order of 1 to 10
nanometers. Such a separation of length scales allows us to study
first a simpler case of two flat surfaces at a distance $w$ from
each other (Fig.~\ref{FigSchematic}b) and only then include
curvature effects.
Each surface carries a dense layer of $\sigma/a^2$ end-grafted
chains per unit area, $a$ being the effective size of a single
monomer. The surfaces are immersed in a melt of mobile polymer
chains, the indices of polymerization of the grafted and free
chains denoted $N$ and $P$, respectively. In the present study
we restrict the discussion to the limit of long free chains
$P\gtrsim N$ and to high grafting densities, while the more
general case will be discussed elsewhere \cite{mmBrush}.
For simplicity we assume that the different monomer species are
all of the same size. We further assume that the different
monomers have the same volume $a^3$, so that the incompressibility
of the polymer melt can be expressed as $\phi_N+\phi_P=1$ where
$\phi_N(x)$ and $\phi_P(x)$ are the local volume fractions of grafted
and free chains, respectively.
The enthalpic interaction between the different monomer
species (in units of the thermal energy $k_BT$) is characterized by
the Flory interaction parameter $\chi$.

\paragraph*{Parabolic Approximation:}

In order to demonstrate the effect of a negative $\chi$ parameter
we implement a variation of the parabolic approximation
\cite{parabolic}, valid in the strong stretching limit of the
grafted chains. In this limit, the mean-field potential felt by a
single chain $\mu_N(x)$ is a quadratic function of the distance
$x$ from the nearest surface \cite{ParabolicRem}. Taking into
account the symmetry around the mid-plane one can write
\begin{eqnarray}\label{muN}
  \beta\mu_N(x) = \left\{
  \begin{array}{ccc}
    A - B x^2     &~   0\le x\le w/2 &\\
    A - B (w-x)^2 &~ w/2\le x \le w  &
  \end{array} \right.
\end{eqnarray} where $B=3\pi^2/8 N^2a^2$, $A(w)$ is an integration
constant to be determined later and $\beta=1/k_BT$. This potential
has to be equal to the sum of the different interactions of the
grafted chain with its environment. The latter can be obtained
from the excess free energy associated with the free chains $V_P =
(1/a^3)\int {\rm d}x v_P(x)$ where \begin{eqnarray}\label{vP}
  \beta v_P &=& {a^2\over 24\phi_P(x)} |\phi_P'(x)|^2
        + {\phi_P(x)\over P}\left[\ln\phi_P-1\right]
        \nonumber\\
        &+& \chi \phi_N(x)\phi_P(x)
\end{eqnarray}
The first term is the elastic energy of the free polymer chains,
the second term is their translational entropy and the last term
is the enthalpic interaction between the grafted and free chains.

Taking into account the incompressibility constraint the mean
field potential can be obtained from the variation of the interaction
energy $\mu_N(x) = -\delta V_P/\delta\phi_N$ which yields
\begin{eqnarray}\label{uN}
     \beta\mu_N(x) =
     ~~~~~ ~~~~~ ~~~~~ ~~~~~ ~~~~~
     ~~~~~ ~~~~~ ~~~~~ ~~~~~ ~~~~~
     \nonumber\\
     - {a^2\over 24 (1-\phi_N(x))}
      \biggl[{1\over 1-\phi_N(x)} |\phi_N'(x)|^2
           + 2\phi_N''(x)\biggr]
      \nonumber\\
      - {1\over P}\ln\left[1-\phi_N(x)\right] -2\chi\phi_N(x)
       ~~~~~ ~~~~~ ~~~~~ ~~~~~ ~~~~~
\end{eqnarray}
Please note that due to the incompressibility of the polymer melt
the last term has the same form as an effective excluded volume
contribution with $v_{\rm eff}\simeq 2|\chi|a^3$.
The above equation combined with eq.~\ref{muN} defines a
differential equation for $\phi_N(x)$ in the interval $0\le x \le
w$, to be solved numerically. Since there is one free parameter,
$A(w)$, three constraints are required. These are provided by mass
conservation, $N\sigma a=\int{\rm d}x\phi_N(x)$, and by two
boundary conditions $\phi_N'=0$ at $x=0$ and $\phi_N'=0$ at
$x=w/2$. The former reflects our choice of a neutral surface
that is equally indifferent to the two polymer species,
while the latter reflects the symmetry of the problem.
The free energy of interaction per unit area can then be
calculated from
\begin{eqnarray}\label{DeltaF}
  F_{\rm pol}(w) &=& {1\over a^3}\int_0^w {\rm d}x
        \left\{v_P(x)-\mu_N(x)\phi_N(x) \right\}
        \nonumber\\
        &+& 2k_BT {\sigma\over a^2} N A(w)
\end{eqnarray}
The above expression results from a transformation from single
chain trajectories to an average volume fraction $\phi_N(x)$. The
first term is the standard free energy of an order parameter in
a local field $\mu_N(x)$ while the last contribution is a reminiscence
of the single chain partition function \cite{parabolic}.

\begin{figure}[tbh]
  \epsfxsize=0.4\linewidth
  \centerline{\hfill\vbox{ \epsffile{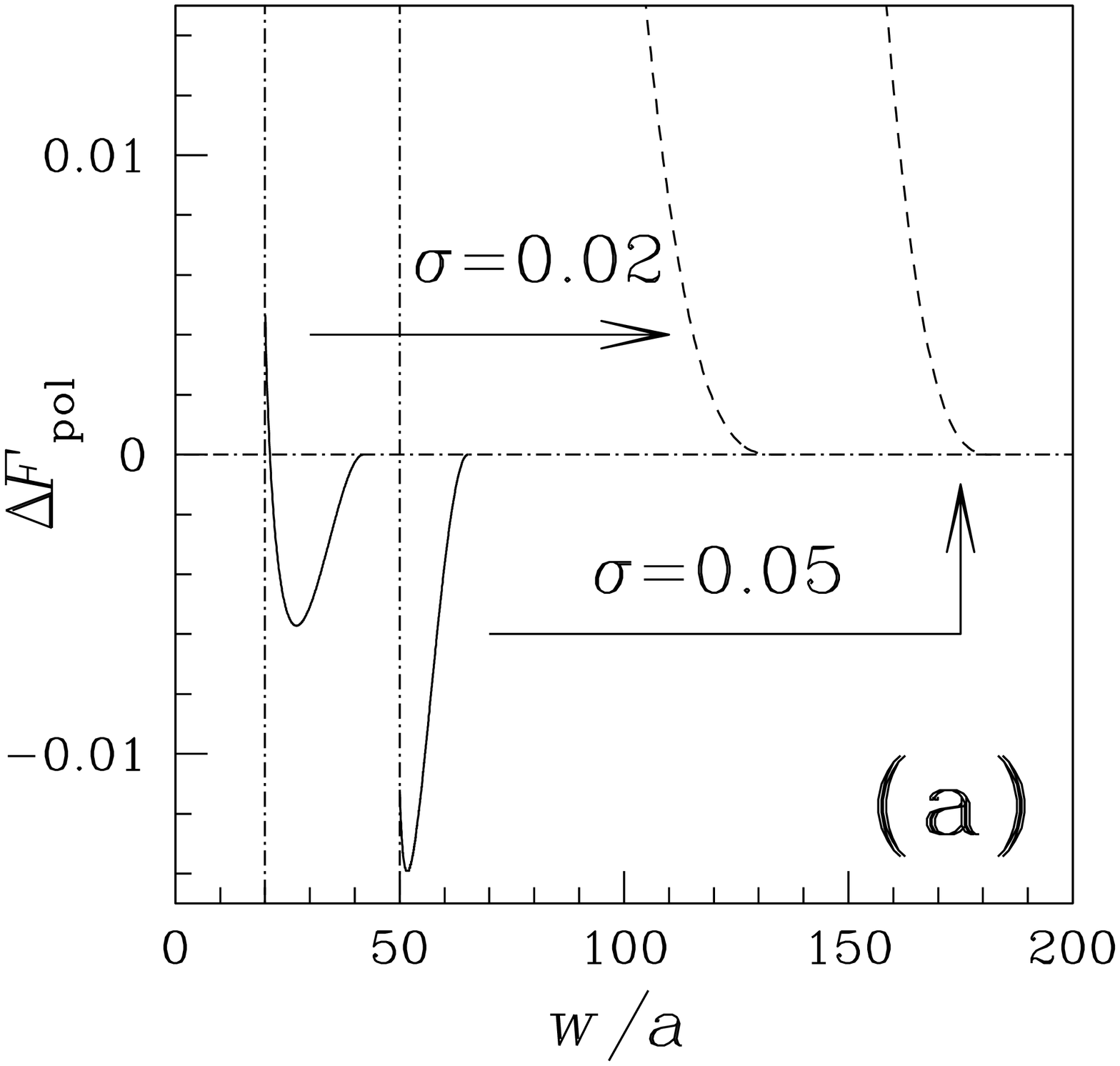}}
  \hfill
  \epsfxsize=0.4\linewidth
              \vbox{ \epsffile{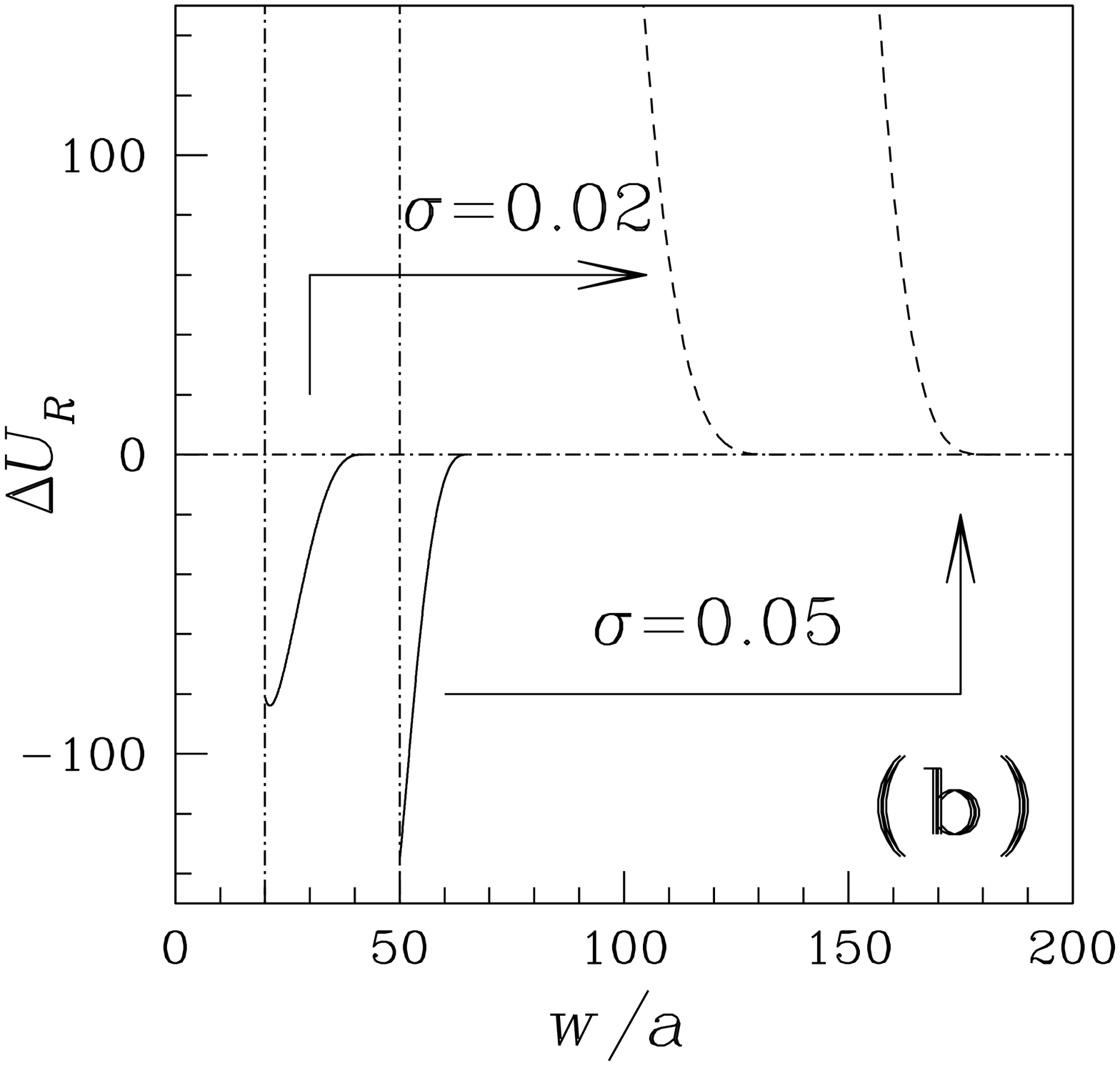}}\hfill}
\vspace{\baselineskip}
\caption{\protect (a) 
Excess free energy of interaction per unit area
$\Delta F_{\rm pol}$ between two flat surfaces (in units of
$k_BT/a^2$) and (b) interaction energy of two grafted colloids
$\Delta U_R$ (in units of $k_BT$) as function of the reduced
inter-surface distance $w/a$. The two solid curves 
correspond to $\chi=0$ with grafting densities $\sigma=0.02$ and
$\sigma=0.05$. The two dashed curves correspond to the same
grafting densities with a negative enthalpic interaction
$\chi=-0.15$. The polymerization indices are $N=P=500$. The radii
of the colloids in (b) is R=400a. 
For clarity, the distances of closest approach ($w=2N\sigma a$) 
are indicated by vertical lines at $w=20a$ and $w=50a$.}
  \label{FigF}
\end{figure}

The effect of a negative $\chi$ parameter is demonstrated in
Fig.~\ref{FigF}a where the excess free energy of interaction
 $ \Delta F_{\rm pol}(w) = F_{\rm pol}(w) - F_{\rm pol}(\infty)$
is plotted as function of the inter-surface separation $w$. When
the two polymer species are chemically the same (namely, $\chi=0$)
an attractive minimum appears at short distances. This attractive
minimum disappears in the presence of attractive enthalpic
interactions between the two species ($\chi=-0.15$ in the figure 
\cite{chi_value}).
This behavior is quite general and below we provide criteria for
the transition between attractive and repulsive behavior.

The interaction energy of two colloids can now be estimated using
the Derjaguin approximation \cite{IsraelachviliBook}
\begin{equation}\label{Derjaguin}
   \Delta U_R(w) = \pi R \int_w^\infty {\rm d}w' \Delta F_{\rm pol}(w')
\end{equation}

    Typical inter-particle interaction energies are plotted in
Fig.~\ref{FigF}b for colloids of radii $R=400a$. Clearly, the
inter--particle attraction disappears when enthalpic interactions
are included.

\begin{figure}[tbh]
  \epsfxsize=0.4\linewidth
  \centerline{\hfill\vbox{\epsffile{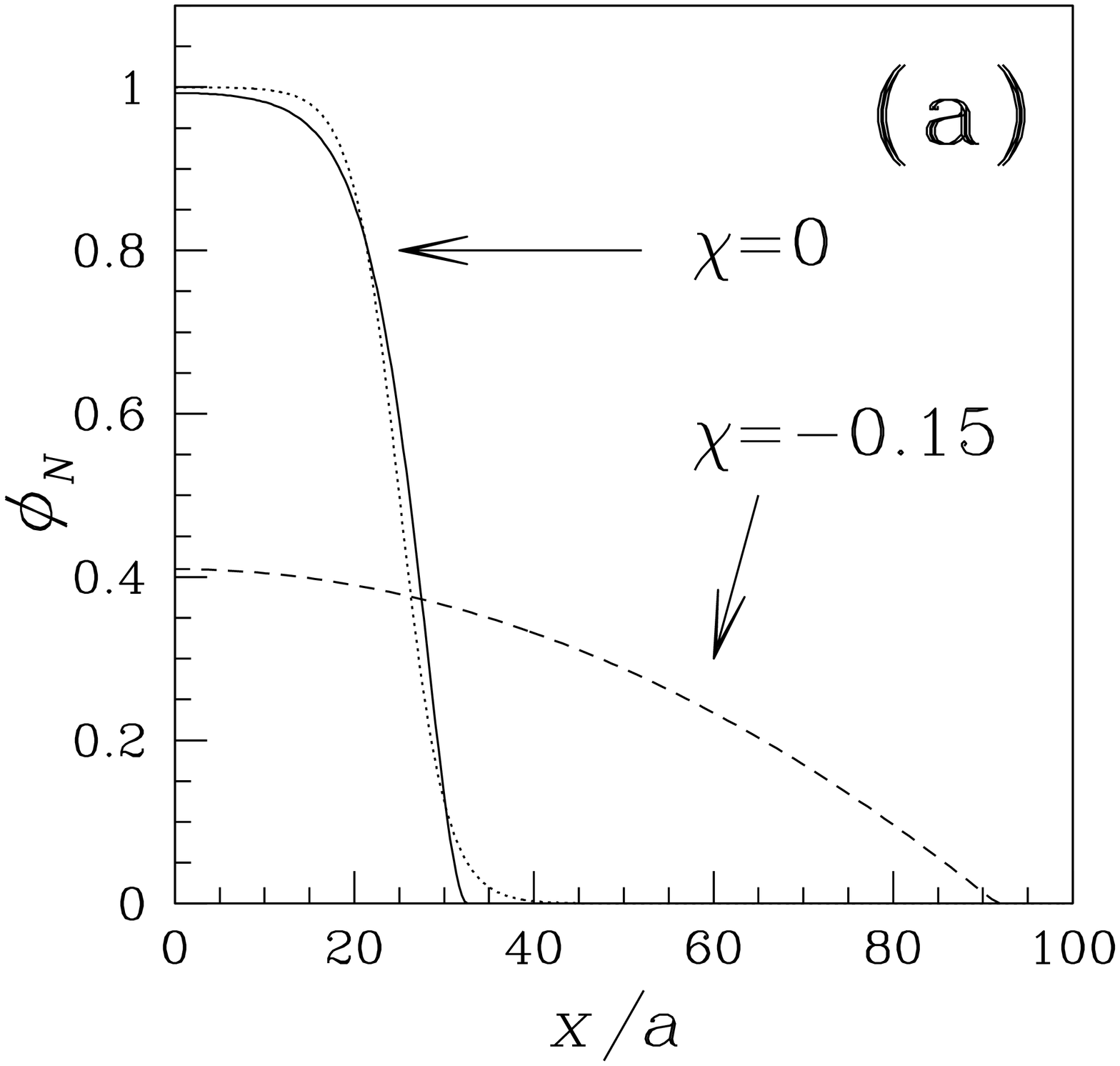}}
  \hfill
  \epsfxsize=0.4\linewidth
              \vbox{\epsffile{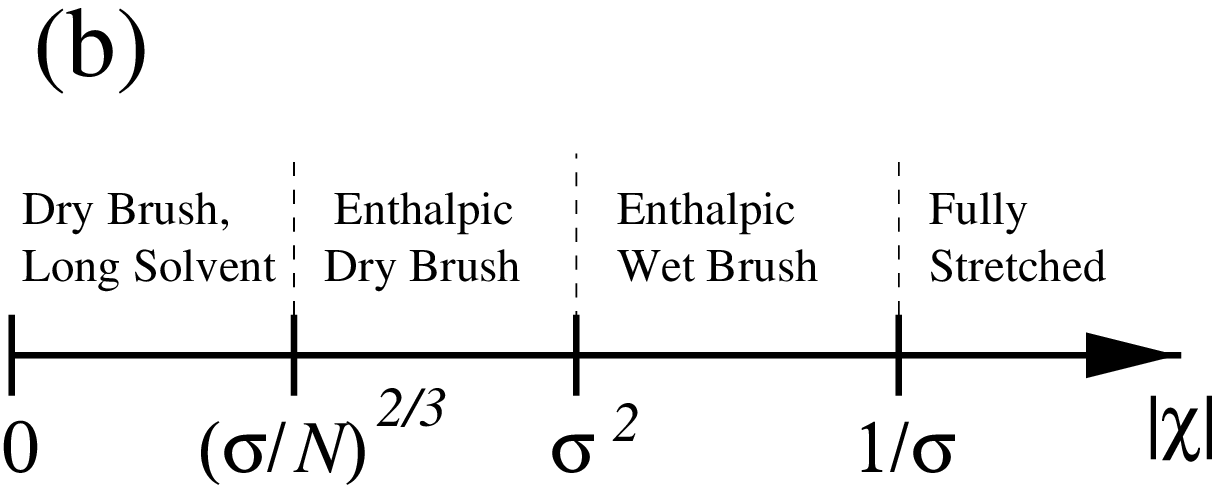}}\hfill}
\vspace{\baselineskip}
\caption{\protect (a) 
Volume fraction profiles $\phi_N$ for $\chi=0$ (solid curve)
and $\chi=-0.15$ (dashed curve)
as function of the distance from the surface $x/a$. The physical
parameters are $\sigma=0.05$ and $N=P=500$. The dotted line
corresponds to the hyperbolic profile with $\lambda=5.15$.
(b) Sequence of brush states as function of the negative
Flory parameter $\chi$. See text for more details
}
  \label{FigDiscussion}
\end{figure}

\paragraph*{Discussion:}
Since the nature of the inter-surface interactions depends
strongly on the shape of the two brushes when they begin to
overlap, it is instructive to consider a single brush in contact
with a polymer melt
\cite{deGennes,Brown,WittenLeibler,Mourran,BrushMelt1}. In
Fig.~\ref{FigDiscussion}a two volume fraction profiles of the
grafted chains are plotted corresponding to $\chi=0$ and
$\chi=-0.15$. Since at high grafting densities the free polymers
are too large to penetrate deep into the grafted layer, the
grafted layer forms a so-called dry brush whose width $h\simeq
N\sigma a$ is determined by packing constraints.

    The free polymers only penetrate a narrow layer of width
$\lambda$, which is determined by the balance between the energy
required to squeeze some free chains into the dense brush and the
elastic response of the polymer chains to local variations in the
density. These can be estimated by assuming a hyperbolic tangent
profile for the volume fraction of the grafted chains
$\phi_N(x)=(1-\tanh[(x-h)/\lambda])/2$. The elastic contribution
to the free energy per unit area is then \cite{Semenov92}
\begin{equation}\label{Fel}
  \beta F_{\rm el} \simeq {1\over 12a\lambda}
\end{equation}
To first approximation the stretching energy can be estimated by
calculating the energy difference between a sharp box-like brush
density and the smooth one \cite{Mourran,Semenov92}:
\begin{equation}\label{Fpen}
  \beta F_{\rm pen} \simeq {\pi^4 \over 32}{\sigma \lambda^2 \over Na^4}
\end{equation}
Minimizing the total free energy $F_{\rm tot}=F_{\rm el}+F_{\rm pen}$
with respect to $\lambda$ yields
\begin{equation}\label{lambda_dry_long}
    \lambda \simeq  \left({4\over 3\pi^4}\right)^{1/3}
                    \left({N\over \sigma}\right)^{1/3} a
\end{equation} For the parameters of Fig.~\ref{FigDiscussion}a one
obtains $\lambda\simeq 5.15a$. The corresponding hyperbolic
profile is indicated in the figure by a dotted line. The good
agreement with the numerical profile demonstrates that indeed
$F_{\rm el}$ and $F_{\rm pen}$ are the dominant terms in the free
energy.

The effect of enthalpic interactions with a negative $\chi$ parameter
on the structure of the brush is somewhat analogous to replacing a
theta solvent with a good solvent where mixing of solvent
molecules with the grafted polymer chains is encouraged \cite{Brown}. The
enthalpic contribution to the free energy of a dry brush can be
easily estimated by
\begin{equation}\label{Fchi}
  \beta F_\chi \simeq {\chi \over a^3} \int{\rm d}x \phi_N(x)\phi_P(x)
    \simeq {\chi\lambda \over 4a^3}
\end{equation} When $|\chi|$ is large enough $F_\chi$ exceeds
$F_{\rm el}$ and the penetration length $\lambda$ results now from
the competition between $F_\chi$ and $F_{\rm pen}$. Thus, for
$|\chi|\gtrsim(\pi^8/48)^{1/3} (\sigma/N)^{2/3}$ the brush enters
the {\em enthalpic dry brush} regime (see diagram in
Fig.~\ref{FigDiscussion}b) where the size of the brush is still
$h\simeq N\sigma a$ but the penetration length follows a different
scaling law \begin{equation}\label{lambda_dry_chi}
  \lambda \simeq {4\over \pi^4} {N|\chi|\over\sigma} a
\end{equation}

   As the enthalpic interactions become stronger, the penetration
length $\lambda$ increases until $\lambda \simeq h$. This happens
for
\begin{equation}\label{dry2wet}
  |\chi|\simeq {\pi^4\over 4}\sigma^2
\end{equation}
At this point the free chains penetrate all the way through the
brush while the grafted chains further stretch into the polymer
melt. The brush is now in the {\em enthalpic wet brush} regime
(see Fig.~\ref{FigDiscussion}b).

    In this regime the enthalpic term dominates the potential
$\mu_N(x)$ (eq.~\ref{vP}) and the density profile is parabolic,
$\phi_N(x)=B(h^2- x^2)/2|\chi|$ while
$h=(8/\pi^2)^{1/3}(|\chi|\sigma)^{1/3}Na$. Indeed, for the
parameters of Fig.~\ref{FigDiscussion}a with $\chi=-0.15$ one
obtains $h\simeq 91a$.

When $|\chi|$ is further increased beyond $|\chi|=\pi^2/8\sigma$,
the grafted chains reach their maximal possible length $h_{\rm
max}=Na$ and we enter the {\em fully stretched brush} regime
(Fig.~\ref{FigDiscussion}b). The parabolic approximation can be
modified in order to take into account the finite extensibility of
the chains \cite{ShimCates} but this is beyond the scope of the
current work.

\paragraph*{Comparison with van der Waals interactions:}
The repulsion between the grafted layers should be now compared
with the van der Waals interactions. These interactions result
from sharp variations in the dielectric properties across an
interface separating two materials. When two surfaces of material
(1) are separated at a distance $l$ by media (2) the non-retarded
van der Waals interaction energy can be written as
\cite{IsraelachviliBook}
\begin{equation}\label{vdW}
  F_{\rm vdW} = - {A_{121}\over 12\pi l^2}
\end{equation}
The Hamaker constant $A_{121}$ depends on the static dielectric
constants of the two media and their refractive indices. When the
surfaces as well as the intervening media are polymers the
contrast is usually not very strong and the Hamaker constants are
typically of order $10^{-21}$J (recall that
$1k_BT\simeq4\times10^{-21}$J).

When two dry brushes interact with each other the contrast at the
brush/melt interfaces should be considered in addition to the
brush/surface interfaces. The former gives a stronger contribution
since the separation between the two brushes $l=w-2N\sigma$ is
smaller than the bare inter-surface separation. In contrast, when
two wet brushes interact with each other the brush/melt interface
is no longer well defined because of the parabolic concentration
profile which decreases rather moderately to zero (see
Fig.~\ref{FigDiscussion}a). Furthermore, since the free melt
chains are present inside the brush the contrast is much smaller.
This leaves only the van der Waals attraction between the bare
surfaces which is too weak at the distance imposed by the swelling
of the brush.

\paragraph*{Conclusions:} In this paper we have suggested a
simple mechanism for stabilizing grafted colloids in a polymer
melt. We have shown that a negative $\chi$ parameter for the
interaction between the grafted polymer chains and the mobile melt
chains will induce swelling of the brush and consequently strong
inter-brush repulsion. This mechanism can have various
implications for creating novel polymer-based materials.

\paragraph*{Acknowledgments:} We would like to thank D. Andelman,
C. Gay, C. Ligoure, H. Orland and P. Pincus for valuable
discussions. I.B. gratefully acknowledges the support of the
French Chateaubriand postdoctoral fellowship and the hospitality
of the Elf-Atochem research center at Levallois-Perret.


\end{multicols}
\vfill

\vfill

\pagebreak
\vfill

\vfill

\vfill

\begin{references}
\bibitem{ParabolicRem}
The parabolic potential can be understood using an analogue from
classical mechanics where the contour of a grafted chain is mapped
onto the trajectory of a point-like particle in an external
potential \cite{parabolic}. The length of the chain is mapped onto
the time it takes the particle to reach the surface. The
constraint that different chain ends are at different distances
from the surface while all chains are of the same length $N$,
means that the time it takes for the particle to reach the surface
is independent of the starting point. This is a basic feature of
an harmonic potential, hence eq.~\ref{muN}.
\end{references}
\end{document}